\documentclass[twocolumn,showpacs,preprintnumbers,amsmath,amssymb]{revtex4}
\usepackage{graphicx}

\def\ket#1{  \left\vert  #1   \right\rangle   }
\def\bra#1{  \left\langle  #1   \right\vert   }

\begin{document}

\title{Quantum Entanglement Percolation}

\author{Michael Siomau}

\email{siomau@nld.ds.mpg.de}

\address{Physics Department, Jazan University, P.O.Box 114,
45142 Jazan, Kingdom of Saudi Arabia and \\ Network Dynamics, Max
Planck Institute for Dynamics and Self-Organization (MPIDS), 37077
G\"{o}ttingen, Germany}

\begin{abstract}
Quantum communication demands efficient distribution of quantum
entanglement across a network of connected partners. The search for
efficient strategies for the entanglement distribution may be based
on percolation theory, which describes evolution of network
connectivity with respect to some network parameters. In this
framework, the probability to establish perfect entanglement between
two remote partners decays exponentially with the distance between
them before the percolation transition point, which unambiguously
defines percolation properties of any classical network or lattice.
Here we introduce quantum networks created with local operations and
classical communication, which exhibit non-classical percolation
transition points leading to the striking communication advantages
over those offered by the corresponding classical networks. We show,
in particular, how to establish perfect entanglement between any two
nodes in the simplest possible network -- the 1D chain -- using
imperfect entangled pairs of qubits.
\end{abstract}


\maketitle

\section{\label{sec:1} Introduction}

Rapid development of quantum technologies implies that large
networks operating with quantum information are to be created in the
nearest future \cite{Kimble:08}. Such quantum networks are superior
to their classical ancestors in security \cite{Gisin:07} and
efficiency of communication \cite{Buhram:01} and exhibit new
structures \cite{Perse:10} and behavior \cite{Manzano:13}, which may
be exploited in quantum computation \cite{Ladd:10} and metrology
\cite{Giov:11}. In general, a network is represented by a graph with
nodes connected by links. In quantum networks, the links are often
associated with quantum channels, through which the nodes exchange
photons \cite{Cirac:97}. The photons may be locally measured with
subsequent classical communication of the results of the
measurements between the nodes.

For efficient communication in a quantum network with certain
configuration of nodes and links, perfect long-distance entanglement
between arbitrary nodes is to be established. As originally
suggested by Acin, Cirac and Lewenstein \cite{Acin:07}, methods of
classical percolation theory \cite{Stauffer:94} can be employed to
find the best strategy for entanglement distribution in a network of
particular configuration leading to the notion of classical
entanglement percolation. Within this notion the communication
capacity of a network is described by its percolation transition
point, before which the probability to establish a path of perfect
entanglement links between two remote nodes is exponentially small,
while this probability has a finite asymptotic limit right after it.
Since each network configuration relates to the transition point,
classical entanglement percolation imposes the fundamental limit on
the communication capacity of the network. This limit can be
overcame by global change of the network configuration with local
operations and classical communication (LOCC) \cite{Acin:07,
Kieling:09, Pers:08, Cuquet:09, Pers:10a, Cuquet:11}. In particular,
by implementing LOCC on a fraction of selected nodes of the network
of initial configuration, such as a honeycomb lattice, one may
redistribute the entanglement between the nodes of a network of
final configuration, e.g. triangular lattice \cite{Acin:07}. This
quantum network reconfiguration reduces the percolation transition
point making quantum communication possible at large scale even with
initially insufficient amount of distributed entanglement. The price
for this advantage is that the selected nodes become disconnected
from the network of final configuration. In other words, the
selected nodes of the initial network must sacrifice their
connectivity for communication benefit of the final network. From
communication viewpoint, this may not be always appreciated at large
scale and especially by the selected nodes. In addition, such a
strategy cannot be implemented on a network of arbitrary initial
configuration. For example, given a triangular lattice or a 1D chain
there is no way to execute the above procedure.

Here we suggest a different approach to overcome the limitations of
the classical entanglement percolation. We employ LOCC to create
complex quantum networks with new percolation properties on networks
of simple initial configuration. Our approach is based on multiple
applications of LOCC and is free of sacrificing nodes of the initial
network. We show, in particular, that the simplest 1D quantum
network with the percolation transition point at unity, can be
modified to the complex so-called hierarchical network
\cite{Boettcher:08, Boettcher:09} with the transition point at $1/2$
using at most polynomial number of LOCC. In the following section we
show first how to transform the 1D chain into the hierarchical
network using LOCC. Then, we give an account on physical resources
required for the transformation in Section~\ref{sec:3} and conclude
in Section~\ref{sec:4}.

\section{\label{sec:2} Entanglement Percolation in 1D chain}

Let us consider a 1D chain, where nodes are placed on a line at
fixed distances from each other and are connected by channels. Two
neighboring nodes may exchange photons through the channels and thus
share pure entangled states of qubits, which may be written in the
computational basis \cite{Nielsen:00} as
\begin{equation}
\label{states}
 \ket{\varphi} = \sqrt{\lambda_1} \ket{00} + \sqrt{\lambda_2} \ket{11}
 \, ,
\end{equation}
where $\lambda_1$ and $\lambda_2$ are the Schmidt coefficients
conditioned by $\lambda_1 \geq \lambda_2$ and $\lambda_1 + \lambda_2
= 1$. A perfect entangled pair with $\lambda_1 = \lambda_2 = 1/2$
can be converted from the above state with the singlet conversion
probability $p = 2 \lambda_2$ by measurement of one of the qubits
from the pair with operators \cite{Vidal:99}
\begin{eqnarray}
  \label{M1-M2}
 M_1 = \left( \begin{array}{cc} \sqrt{\frac{\lambda_2}{\lambda_1}} & 0 \\ 0 & 1
\end{array} \right) , \;
 M_2 = \left( \begin{array}{cc} \sqrt{1 - \frac{\lambda_2}{\lambda_1}} & 0 \\ 0 &
0 \end{array} \right) \, .
\end{eqnarray}

In quantum communication the singlet conversion probability plays
the exact role of link occupation probability in percolation theory.
Indeed, if the imperfect entangled pairs are distributed between the
nodes of a quantum network, each of the pairs is converted to the
perfect entanglement link with probability $p$ or vanish with
probability $1-p$. By analogy with the percolation theory, we say
that perfect entanglement is established between two remote nodes if
there is a path of connected perfect entanglement links between the
nodes. While the singlet conversion probability is a natural choice
to study the entanglement percolation, we shall also use another
measure of entanglement to characterize the process of entanglement
distribution -- the concurrence \cite{Wootters:98}. The concurrence
for the entangled qubit pair (\ref{states}) is given by $\mathcal{C}
= 2 \sqrt{\lambda_1 \lambda_2}$. Perfect entanglement with
concurrence $\mathcal{C}=1$ is established between two distant nodes
if only there is a path of connected entanglement links of
concurrence $\mathcal{C}=1$ each.

\begin{figure}
\begin{center}
\includegraphics[scale=0.3]{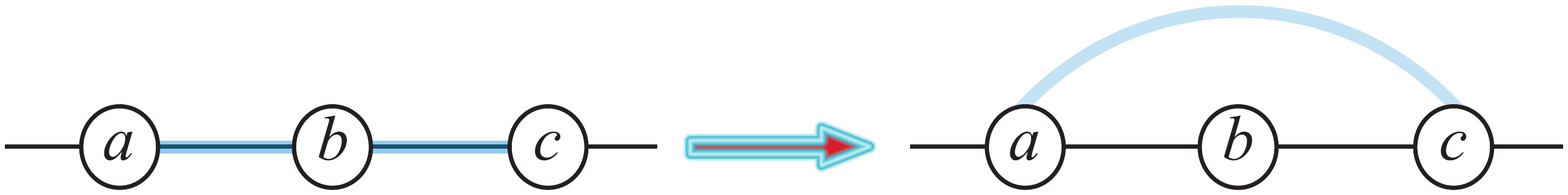}
\caption{Entanglement swapping at node $b$ leads to the creation of
a non-local entanglement link connecting two physically disconnected
nodes.}
 \label{fig-1}
\end{center}
\end{figure}

The main ingredient for the  advanced quantum information processing
is the entanglement swapping \cite{Pan:98}. If three nodes $a-b-c$
are chained together and the node $b$ shares two qubit pairs in
state (\ref{states}) with the neighbors, the optimal entanglement
swapping \cite{Pers:08} consists of the measurement of the two
qubits from the two different entangled pairs at node $b$ in the
Bell basis
\begin{eqnarray}
 \label{Bell-measurements}
\ket{\Psi^{\pm}} &=& \frac{1}{\sqrt{2}} \left( \ket{00} \pm \ket{11}
\right) \, , \nonumber
 \\[0.1cm]
\ket{\Phi^{\pm}} &=& \frac{1}{\sqrt{2}} \left( \ket{01} \pm \ket{10}
\right) \, ,
\end{eqnarray}
with subsequent classical communication of the results of the
measurement to nodes $a$ and $c$ (see Fig.~\ref{fig-1}). The two
local entanglement links transform into a single non-local
entanglement link with the final state given by
\begin{eqnarray}
 \label{Final-Swapping-1}
\ket{\psi^{\pm}} &=& \frac{\lambda_1}{\sqrt{\lambda_1^2 +
\lambda_2^2}} \ket{00} \pm  \frac{\lambda_2}{\sqrt{\lambda_1^2 +
\lambda_2^2}} \ket{11} \, ,
 \\[0.1cm]
 \label{Final-Swapping-2}
\ket{\phi^{\pm}} &=& \frac{1}{\sqrt{2}} \left( \ket{01} \pm \ket{10}
\right) \, ,
\end{eqnarray}
with corresponding probabilities $(\lambda_1^2 + \lambda_2^2)/2$ and
$\lambda_1 \lambda_2$. The entanglement swapping doesn't change the
average singlet conversion probability $p_{\rm swap} = p$
\cite{Bose:99} and reduce concurrence $\mathcal{C}_{\rm swap} = 2
\lambda_1 \lambda_2 \equiv \alpha \mathcal{C}$ \cite{Pers:08}, where
$\alpha= \sqrt{\lambda_1 \lambda_2}$. The possibility to use LOCC to
create a non-local entanglement link, which is beyond the initial
network configuration, allows us to create a complex quantum network
with highly unexpected large-scale behavior as we show next.

\begin{figure}
\begin{center}
\includegraphics[scale=0.3]{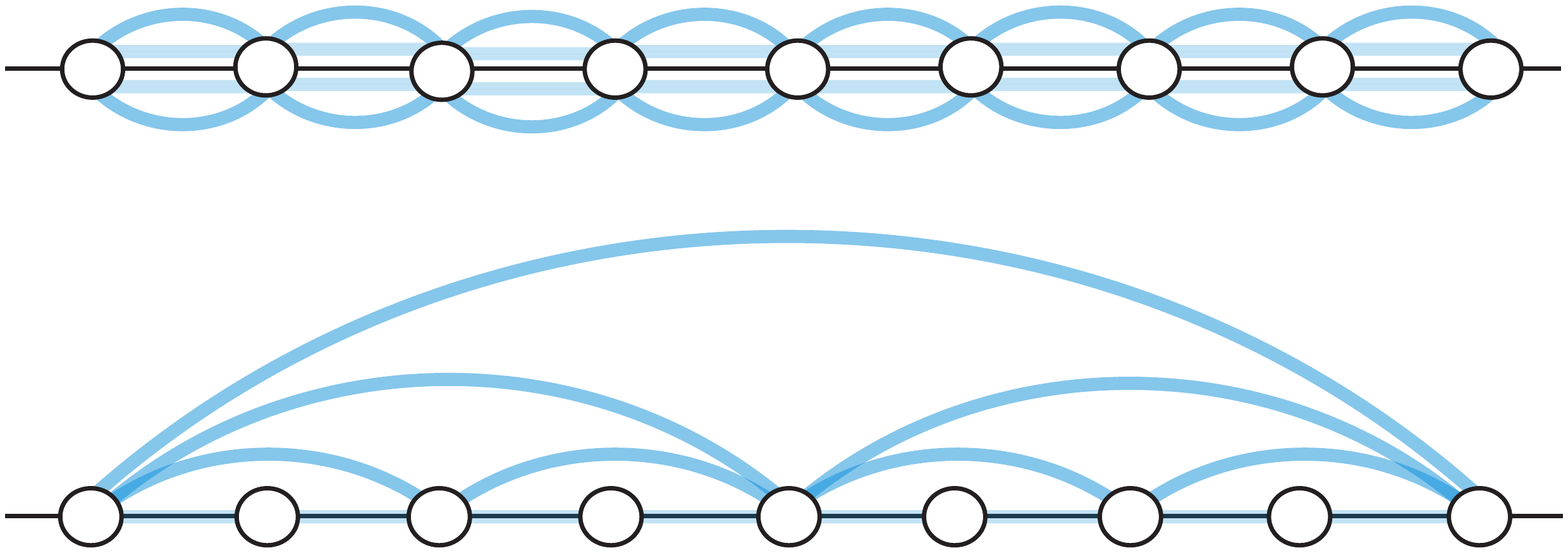}
\caption{The 1D chain of $9$ nodes with $4$ entangled pairs between
neighboring nodes (top) is modified with the entanglement swapping
at each node - ignoring the borders - to the quantum network with
$3$ levels of hierarchy (bottom).}
 \label{fig-2}
\end{center}
\end{figure}

Let us consider a 1D chain consisting of $N$ nodes. Let each pair of
neighboring nodes initially share $K$ qubit pairs in the state
(\ref{states}). Let us take $K-1$ qubit pairs and perform
entanglement swapping at each second node. After that we take $K-2$
qubit pairs and apply the entanglement swapping at each third node.
Repeating the above procedure with the entanglement swapping at
$2^{k}+1$ nodes for $k=0...K-2$, we construct $K-1$ levels of
hierarchy of non-local entanglement links over the initial 1D chain
formed with local links, as exemplified in Fig.~\ref{fig-2}. This
hierarchical network was introduced by Boettcher, Goncalves and
Guclu \cite{Boettcher:08} to study the behavior of the so-called
small-world systems \cite{Watts:98}. But, because the non-local
entanglement links are created with LOCC only and do not require any
non-local interactions between the nodes, they cannot be attributed
to the small-world links in a common classical sense. Moreover, the
non-local links don't correspond to physical communication channels:
the physical connectivity of the chain is still one-dimensional.
Thus, created hierarchical network of entanglement links has no
classical analog and may be called a genuine quantum network
constructed on a 1D chain, while the entanglement percolation in
such a network -- the quantum entanglement percolation. The
suggested approach demands creating non-local entanglement links
beyond the given network configuration and cannot be implemented
classically.

Now, we show that quantum entanglement percolation in the
hierarchical quantum network is of exponential benefit over the
entanglement percolation in 1D chain. In the 1D chain with $K$
entangled pairs between to neighboring nodes, perfect entanglement
between two neighboring nodes is established with probability
$p^{\prime} = p + p (1 - p)^K$. The condition for connecting two
infinitely distant nodes with a perfect entanglement link reads as
\begin{equation}
\lim_{N\rightarrow \infty} (p^{\prime})^N = 1 \, ,
\end{equation}
where $N$ is the number of nodes between the nodes to be connected.
The condition holds true only for $ p^{\prime} = p = 1$ giving us
the well-known percolation transition point in the classical 1D
chain $p_t = 1$, i.e. in 1D configuration the perfect entanglement
can be created between two infinitely distant nodes if only the
perfect entanglement is established between each pair of neighboring
nodes.

For the quantum hierarchical network, the probability to connect two
neighboring nodes is $P_0 = p$, while for three nodes $P_{1} = p +
(1-p) P_0^2$, where we used the fact that the entanglement swapping
doesn't change the singlet conversion probability. Taking into
account self-similarity of the chain, we arrive to the recursive
formula to find the probability of connecting the border nodes with
a path of perfect entangled pairs \cite{Boettcher:09}
\begin{eqnarray}
 \label{recursion}
P_{k+1} = p + (1-p) P_k^2 \, ,
\end{eqnarray}
for $k=0...K-2$. In the infinite chain $N \rightarrow \infty$, thus
$K \rightarrow \infty$ and $P_{\infty} = P_{K-1} \approx P_{K-2}$.
The quadratic equation has two solutions interconnected at the
non-classical percolation transition point $p_t=1/2$. It is
important to stress that, in contrast to standard percolation
\cite{Stauffer:94}, before the percolation transition point $p <
1/2$ there is a finite probability of connecting the border nodes
$P_{\infty} = p/(1-p)$. However, further analysis
\cite{Boettcher:12} shows, that the probability to connect an
arbitrary two nodes of the network with a path of perfect
entanglement links is still exponentially small due to exponentially
small size of the giant component with respect to the network size.
For $p > 1/2$ a path between arbitrary two nodes in the network
exists with a finite probability \cite{Boettcher:12}. Thus the
initially distributed entangled states (\ref{states}) in the
hierarchical network with the singlet conversion probability of $p
\geq 1/2$ are sufficient to establish a path of perfect entanglement
links between arbitrary two nodes irrespective of the chain length.
Taking into account that in the 1D chain the perfect entanglement
path exists only if $p=1$ (and otherwise $P_{\infty} = 0$), for any
$p > 1/2$ quantum entanglement percolation is superior exponentially
to the classical entanglement percolation as was announced.

In Eq.~(\ref{recursion}) we implicitly assumed that the average
singlet conversion probability doesn't change after multiple
entanglement swapping operations. But, this is not the case, because
after a single entanglement swapping with the Bell measurements
(\ref{Bell-measurements}), the pure state (\ref{Bell-measurements})
transforms into a mixture of states (\ref{Final-Swapping-1}) and
(\ref{Final-Swapping-2}). Further application of the entanglement
swapping on the mixed state leads to the decrease of the average
singlet conversion probability. Since the entanglement swapping is a
LOCC and the entanglement doesn't increase under LOCC, there must be
an optimal measurement that preserves the entanglement under
multiple application of the entanglement swapping. But, as the
optimal measurement is unknown yet \cite{Pers:08}, in the next
section we shall estimate the amount of initial entangled pairs
(\ref{states}) and number of simple entanglement swapping operations
with the Bell measurements (\ref{Bell-measurements}) to create the
hierarchical network on the 1D chain.

\section{\label{sec:3} Properties of the Entanglement Percolation}

Before we proceed with the percolation properties of the quantum
network, we would like to stress that each of the $K-1$ levels of
hierarchy connects $2^{K-1}+1$ nodes into loops with one entangled
link between the border nodes. Thus to construct a network of
maximal hierarchy on $N$ nodes (i.e. to connect the border nodes
with a single entanglement link) one needs to distribute initially
at least $ 1 + \log_2 (N-1)$ imperfect entangled pairs between each
pair of neighboring nodes in the initial 1D chain. The total number
of the imperfect entangled states in the hierarchical network of $N$
nodes thus scales as $N \left( 1 + \log_2 (N-1)\right) \ll N^2$,
which is practically feasible.

Studying the quantum entanglement percolation with concurrence gives
us further insight into the process of entanglement distribution in
the quantum network. Because single entanglement swapping reduces
the concurrence of the initial states $\alpha < 1$ times, the
recursion relation (\ref{recursion}) is modified as
\begin{equation}
P_{k+1} = \alpha^{k+1} \mathcal{C} + (1- \alpha^{k+1} \mathcal{C})
P_k^2 \, ,
\end{equation}
where $P_0 = \mathcal{C} = 2 \sqrt{\lambda_1 \lambda_2}$. Because
$\alpha^{k+1} \mathcal{C} \rightarrow 0$ as $k \rightarrow \infty$,
the percolation properties of the hierarchical network reduce to the
percolation properties of the classical 1D chain $P_{\infty} =
P_{K-1} \approx P_{K-2}^2 \propto \mathcal{C}^N$ with the classical
percolation transition point at $\mathcal{C}_t = 1$. The reason for
the reduction is the decay of entanglement in the non-local
entanglement links of higher hierarchy due to multiple entanglement
swapping. The decay is exponential with respect to the hierarchy
level $K-1$, which implies polynomial decay of entanglement with
respect to the chain length $N$, because $\alpha \leq 1/2$ and $N
\propto 2^{K-1}$.

The effect of the polynomial decay of entanglement can be eliminated
using standard protocol for entanglement distillation
\cite{Bennet:96}. Reminding that any two-qubit state is distillable
if entangled \cite{Horodecki:09}, let us estimate the amount of
initial entanglement to implement the distillation in the
hierarchical network. Let us consider the initial network
configuration with $K$ entangled pairs between the neighboring nodes
and assuming that all qubit pairs (\ref{states}) have the amount of
entanglement of $\mathcal{C} = 1/2$, which corresponds to the states
with Schmidt coefficients $\lambda_{1,2} =1/2 \pm \sqrt{3}/4$. These
states could be unitary transformed into Werner states
\cite{Horodecki:09} with fidelity of the each of the states $F
\equiv {\rm Tr} \left( \ket{\varphi}\bra{\varphi} \ket{\Psi^{+}}
\bra{\Psi^+} \right) = \left( 1 + 2 \sqrt{\lambda_1 \lambda_2}
\right)/2 = 3/4$. A single entanglement swapping at an arbitrary
node of the 1D chain results into one of the two states
(\ref{Final-Swapping-1})-(\ref{Final-Swapping-2}). While the state
(\ref{Final-Swapping-2}) requires no distillation, the fidelity of
the state (\ref{Final-Swapping-1}) is given by $F_{(0)} = 1/2 +
1/(4\sqrt{14}) \approx 0.57$. Using the iterative formula for the
entanglement distillation
\begin{eqnarray}
F_{(i+1)} = \frac{F_{(i)}^2 + \frac{1}{9}(1-F_{(i)})^2}{F_{(i)}^2 +
\frac{2}{3} F_{(i)} (1-F_{(i)}) + \frac{5}{9}(1-F_{(i)})^2} \, ,
\end{eqnarray}
we find that $F_{(0)} \rightarrow F_{(8)} > F$, i.e. the
entanglement distillation protocol allows to restore the fidelity to
the level before the entanglement swapping in just eight iterations.
Taking into account that the success probability of the distillation
protocol approximates $1/4$, $32$ entangled pairs are required to
distill a single entanglement link after an entanglement swapping.
In a network with $K-1$ levels of hierarchy, the number of entangled
pairs before distillation thus scales exponentially as $ 2^{K-2}
\times 32^{K-1} \propto 2^{6 K}$, but because $K \propto \log_2 N$,
the total number of initial states to construct the hierarchical
network with the average concurrence $\mathcal{C} \geq 1/2$ per
entanglement link scales just polynomially $\propto N^6$. The
distillation procedure is efficient, because targets achieving
states with non-unit fidelity $F \geq 3/4$ at each entanglement
link. If each entanglement link in the hierarchical quantum network
has the average concurrence $\mathcal{C} \geq 1/2$, the probability
to establish a perfect entanglement path between two arbitrary
distant nodes is strictly higher then zero as suggested by the
percolation properties of the quantum network.

Suggested distillation procedure is not optimal and is shown to
demonstrate practically appropriate (i.e. polynomial) scaling of the
initial resources with the network size. More advanced distillation
protocols \cite{Horodecki:09} may lead to even better scaling. The
optimal strategy for the entanglement distillation depends on the
optimal strategy for the hierarchical network construction, which is
unknown yet, as we mentioned early.

\section{\label{sec:4} Conclusion}

We introduced quantum networks created with LOCC, which exhibit
non-classical percolation properties and have quantum communication
advantages over corresponding classical networks. Using the notion
of quantum entanglement percolation we showed how to establish
long-distance perfect entanglement between arbitrary two nodes in 1D
chain with imperfect entangled pairs. Apart from the quantum
communication benefits, we clearly demonstrated the distinction
between the percolation properties of the physical network
configuration (composed of nodes and channels) and the quantum
network configuration (consisting of nodes and entanglement links).
We showed that the percolation properties of these networks are
dramatically different, although they both correspond to the
physical 1D configuration. This result suggests a study of
structural complexity of entanglement graphs that can be simulated
on a given quantum network \cite{Siomau:16}.

Presented approach of constructing the non-local hierarchical levels
on simple underlying classical networks can be extended beyond 1D
network configuration \cite{Boettcher:09}. In particular, the 2D
square lattice with the classical percolation transition point at
$p=1/2$ can be modified to a hierarchical network with the
percolation transition point at $p=5/32 \approx 0.16$. General
analysis of the hierarchical networks is, however, challenging and
requires development of new theoretical and numerical tools.

The hierarchical networks exhibit property of explosive percolation
\cite{Boettcher:12} -- the sudden emergence of large-scale
connectivity in a network \cite{Souza:15}. The fact that the
hierarchical networks may be created and operated locally opens
intriguing possibilities for experimental testing of the explosive
percolation.

Finally because the quantum networks exhibit new percolation
transition points, we may expect all-new percolation properties,
which may lead to the construction of new local theory of
percolation in quantum networks with new unexpected technological
applications.

\begin{acknowledgments}
I thank Marc Timme for hospitality in his group and Malte
Schr\"{o}der for stimulating discussions. This work was supported by
KACST under research grant 180-35.
\end{acknowledgments}


\begin{thebibliography}{30}

\bibitem{Kimble:08}
H.J.~Kimble, Nature \textbf{453}, 1023 (2008).

\bibitem{Gisin:07}
N.~Gisin and R.~Thew, Nature Photon. \textbf{1}, 165 (2007).

\bibitem{Buhram:01}
H.~Buhram, R.~Cleve, J.~Watrous and R.~de Wolf, Phys. Rev. Lett.
\textbf{87}, 167902 (2001).

\bibitem{Perse:10}
S.~Perseguers, M.~Lewenstein, A.~Acin and J.I.~Cirac, Nature Phys.
\textbf{6}, 539 (2010).

\bibitem{Manzano:13}
G.~Manzano, F.~Galve, G.L.~Giorgi, E.~Hernandez-Garcya and
R.~Zambrini, Scientific Reports \textbf{3}, 1439 (2013).

\bibitem{Ladd:10}
T.D.~Ladd, F.~Jelezko, R.~Laflamme, Y.~Nakamura, C.~Monroe and
J.L.~O'Brien, Nature \textbf{464}, 45 (2010).

\bibitem{Giov:11}
V.~Giovannetti, S.~Lloyd and L.~Maccone, Nature Photon. \textbf{5},
222 (2011).

\bibitem{Cirac:97}
J.I.~Cirac, P.~Zoller, H.J.~Kimble and H.~Mabuchi, Phys. Rev. Lett.
\textbf{78}, 3221 (1997).

\bibitem{Acin:07}
A.~Acin, J.I.~Cirac and M.~Lewenstein, Nature Phys. \textbf{3}, 256
(2007).

\bibitem{Stauffer:94}
D.~Stauffer and A.~Aharony, \textit{Introduction to Percolation
Theory} (Taylor \& Francis, Ed. 2, 1994).

\bibitem{Kieling:09}
K.~Kieling and J.~Eisert, Lect. Notes Phys. \textbf{762}, 287
(2009).

\bibitem{Pers:08}
S.~Perseguers, J.I.~Cirac, A.~Acin, M.~Lewenstein and J.~Wehr, Phys.
Rev. A \textbf{77}, 022308 (2008).

\bibitem{Cuquet:09}
M.~Cuquet and J.~Calsamiglia, Phys. Rev. Lett. \textbf{103}, 240503
(2009).

\bibitem{Pers:10a}
S.~Perseguers, D.~Cavalcanti, G.J.~Lapeyre Jr., M.~Lewenstein and
A.~Acin, Phys. Rev. A \textbf{81}, 032327 (2010).

\bibitem{Cuquet:11}
M.~Cuquet and J.~Calsamiglia, Phys. Rev. A \textbf{83}, 032319
(2011).

\bibitem{Boettcher:08}
S.~Boettcher, B.~Goncalves and H.~Guclu, J. Phys. A.: Math. Theor.
\textbf{41}, 252001 (2008).

\bibitem{Boettcher:09}
S.~Boettcher, J.L.~Cook and R.M.~Ziff, Phys. Rev. E \textbf{80},
041115 (2009).

\bibitem{Nielsen:00}
M.A.~Nielsen and I.J.~Chuang, \textit{Quantum Computation and
Quantum Information} (Cambridge Univ. Press, Cambridge, 2000).

\bibitem{Vidal:99}
G.~Vidal, Phys. Rev. Lett. \textbf{83}, 1046 (1999).

\bibitem{Wootters:98}
W.K.~Wootters, Phys. Rev. Lett. \textbf{80}, 2245 (1998).

\bibitem{Pan:98}
J.-W.~Pan, D.~Bouwmeester, H.~Weinfurter and A.~Zeilinger, Phys.
Rev. Lett. \textbf{80}, 3891 (1998).

\bibitem{Bose:99}
S.~Bose, V.~Vedral and P.I.~Knight, Phys. Rev. A \textbf{60}, 194
(1999).

\bibitem{Watts:98}
D.J.~Watts and S.H.~Strogatz, Nature \textbf{393}, 440 (1998).

\bibitem{Boettcher:12}
S.~Boettcher, V.~Singh and R.M.~Ziff, Nature Commun. \textbf{3}, 787
(2012).

\bibitem{Bennet:96}
C.H.~Bennett, G.~Brassard, S.~Popescu, B.~Schumacher, J.A.~Smolin,
and W.K.~Wooters, Phys. Rev. Lett. \textbf{76}, 722 (1996).

\bibitem{Horodecki:09}
R.~Horodecki, P.~Horodecki, M.~Horodecki and K.~Horodecki, Rev. Mod.
Phys. \textbf{81}, 865 (2009).

\bibitem{Siomau:16}
M.~Siomau, AIP Conf. Proc. \textbf{1742}, 030017 (2016).

\bibitem{Souza:15}
R.M.~D'Souza and J.~Nagler, Nature Phys. \textbf{11}, 531 (2015).

\end{thebibliography}
\end{document}